\documentclass{article}
\usepackage{amsmath,cases,amssymb,mathrsfs,color,braket,here,ascmac,enumerate,multirow}
\usepackage[dvipdfmx]{graphicx}
\usepackage{mathtools,array,booktabs,colortbl}

\usepackage[margin=25truemm]{geometry}

\mathtoolsset{showonlyrefs=true}

\makeatletter

\def\affiliation#1{\def\@affiliation{#1}}

\def\@maketitle{
\begin{center}
{\Large \bf \@title \par}
\vspace{5mm}
{\large \@author \par}
{\normalsize \it \@affiliation \par}
\end{center}
}

\newcolumntype{C}[1]{>{\hfil}m{#1}<{\hfil}}

\@addtoreset{equation}{section}

\makeatother

\title{No-Go Theorems for Hairy Black Holes in Scalar- or Vector-Tensor-Gauss-Bonnet Theory}
\author{Satoshi Matsumoto}
\affiliation{Department of Physics, Kindai University, Higashi-Osaka, Osaka 577-8502, Japan}

\begin{document}

\maketitle

\begin{abstract}
In this paper, we show a no-go theorem for static spherically symmetric black holes with vector hair in Einstein-$\Lambda$-Vector-Tensor-Gauss-Bonnet theory where a complex vector field non-minimally couples with Gauss-Bonnet invariant. For this purpose, we expand metric functions and radial functions of a vector field around the event horizon, and substitute the expansions into equations of motion. Demanding that the equations of motion are satisfied in each order, we show that the complex vector field vanishes on the event horizon. Moreover, when the event horizon is degenerated, it is also implied that the complex vector field vanishes on and outside the horizon. In addition, we study the case in which the vector field non-minimally couples extra, and examine the no-hair theorem with different coupling functions.
\end{abstract}

\section{Introduction}
\label{intro}

The statement that isolated black hole solutions are characterized entirely by a mass, electric charge, and angular momentum is called no-hair theorem by likening above three physical quantities to hairs. Almost all astrophysical objects existing in the universe are characterized by a large amount of physical quantities. Therefore the black holes that are characterized by at most three quantities are simple objects. In general relativity, for stationary, axially symmetric spacetime, Bekenstein proved \cite{bekenstein1, bekenstein2} that in the formation of a black hole, matter fields described by scalar and vector fields except for massless charged vector fields interpreted as electro-magnetic fields, are radiated away or fall into the event horizon, and hence the black hole has no hair. In the actual universe, however, there are accretions of a variety of matters (e.g., scalar and vector fields) from stellar gases into a black hole. Thus, it may be not ensured that the no-hair theorem is held in more general situations. There are a number of papers which examine the possibilities of hairy black holes \cite{gibbons}-\cite{corelli2}.

The hair of a black hole can be made from either an undiscovered matter or some known matter in the standard model. When we consider the hair as an undiscovered matter, cosmologically interesting possibilities are a candidate of dark sector (e.g., dark matter or dark energy). It is important to consider how we can distinguish hairy black holes and bald ones through astrophysical observations such as detection of gravitational waves from black holes binaries. It is also important to study what type of matter fields can play a role of a black hole hair.

One of the interesting methods to construct hairy black holes is to consider gravitational theories which include non-minimal couplings between matter fields and source terms. In such a method, various hairy black hole solutions were discovered. Gibbons et al. \cite{gibbons} derived exact solutions of a black hole and a membrane for general dimensions in the theory where a dilaton non-minimally couples with an anti-symmetric 2-form. In 4-dimensions, Garfinkle et al. \cite{garfinkle} independently derived an exact solution of the hairy black hole with electric fields. Herdeiro et al. \cite{herdeiro1} numerically constructed the hairy black hole solution in Einstein-Maxwell-Scalar (EMS) model where a scalar field non-minimally coupled with a Maxwell source term. In EMS model, it was shown that black holes can be spontaneously scalarized due to tachyonic instability arisen from a quadratic coupling function. Fernandes et al. \cite{fernandes} varied the coupling function and investigated corresponding changes and dynamical features of the solution. Brihaye et al. \cite{brihaye} numerically constructed hairy black hole solutions and their spontaneous scalarizations in EMS-$\Lambda$ model and extended Scalar-Tensor-Gauss-Bonnet-$\Lambda$ (STGB-$\Lambda$) model. EMS-$\Lambda$ model is a gravitational theory including a positive cosmological constant and a scalar field non-minimally coupled with a Maxwell source term, while extended STGB-$\Lambda$ model is a theory including a positive cosmological constant and a scalar field non-minimally coupled with a GB invariant. Elley et al. \cite{elley} studied spontaneous scalarization and descalarization of black hole binaries. Corelli et al. \cite{corelli1, corelli2} examined effects of higher order curvature on black hole evaporation in STGB theory with a minimally coupled phantom scalar field. Furthermore, in general relativity,  Santos et al. \cite{santos} considered black hole with Proca hair by superradiance. Their result does not conflict with Bekenstein's no-hair theorem \cite{bekenstein1, bekenstein2}, since their assumption on the symmetry of Proca fields and background geometry is different from that assumed in Bekenstein's work. Therefore, their result \cite{santos} suggests the possibility of hairy black holes in general relativity.

As mentioned above, the hairy black holes obtained so far are mostly numerical solutions. Even if one simplifies the equations of motion by imposing symmetries, it is still very hard to find exact hairy solutions analytically. The purpose of this paper is to analytically show a no-go theorem for hairy black holes with complex vector fields in Vector-Tensor-Gauss-Bonnet (VTGB) theory with a non-minimal coupling, by applying a method developed in STGB theory with a non-minimal coupling between a complex scalar field and GB invariant. Considering complex scalar and vector fields around a black hole offers the interesting possibility of constructing a less symmetric black hole which could describe a final state of unstable rotating black hole, as demonstrated in \cite{dias}.

We also mention the following remarks. In VTGB theory, when GB term is ignored, the VTGB theory corresponds to general relativity. Moreover, we demand that the coupling function become a constant when the vector fields vanish, so that the VTGB theory reduces to the Einstein-Gauss-Bonnet (EGB) theory whose equations of motion are consistent with those in general relativity.

This paper is organized as follows. First, in section \ref{nogo_scalar}, we analyze the no-go theorem for an extremal black hole with complex scalar fields as an extension of previous work \cite{lin}, and confirm that the no-go theorem is held also in the extremal case. Next, in  section \ref{nogo_vector}, we show the no-go theorem for both extremal and non-extremal black holes with complex vector fields. Furthermore, in section \ref{EC_CD}, we examine the no-hair theorem in the theories in which the matter field couples with Ricci scalar and the cosmological constant in addition to the GB invariant, and with different coupling functions. Finally, in section \ref{concl}, we summarize all results obtained by our analyses.

\section{No-go theorem for black holes with the complex scalar hair}
\label{nogo_scalar}

In this section, we show that in the $\Lambda$-STGB theory, a spherically symmetric, static black hole with a degenerate event horizon cannot have a complex scalar hair. This no-go theorem is an extension of the previous work \cite{lin} to the extremal black hole case.

We consider the scalar-tensor-Gauss-Bonnet gravity with a cosmological constant $\Lambda$ described by the action
\begin{equation}
\label{action_scalar}
\mathcal{S} = \int d^4x \sqrt{-g} \left[ R - 2\Lambda + c_1 f(\eta) \mathcal{G} + c_2 \nabla_\mu \Phi \nabla^\mu \bar{\Phi} + c_3 V(\eta) \right],
\end{equation}
where $R$ is the Ricci scalar, $\Lambda$ is the cosmological constant, and
\begin{equation}
\label{gaussbonnet}
\mathcal{G} \equiv R^2 - 4R_{\alpha \beta} R^{\alpha \beta} + R_{\alpha \beta \gamma \delta} R^{\alpha \beta \gamma \delta}
\end{equation}
is the Gauss-Bonnet invariant with $R_{\mu \nu}$ and $R_{\alpha \beta \gamma \delta}$ being the Ricci tensor and Riemann tensor, respectively. Here $\Phi$ is a complex scalar field, $f(\eta)$ and $V(\eta)$ are respectively the non-minimal coupling function and the potential which are  arbitrary functions of $\eta \equiv \Phi \bar{\Phi}$. $c_i$ ( $i = 1, 2, \cdots$ ) are real constants. In this theory, we will establish the following theorem: {\it In general Einstein-$\Lambda$-STGB systems, static spherically symmetric black holes cannot have a complex scalar hair. Furthermore, complex scalar fields vanish over the spacetime (i.e., on and outside the event horizon), when the event horizon is degenerated}.

Through this paper, we consider the static spherically symmetric spacetime whose metric is, in general, described by
\begin{equation}
\label{metric}
ds^2 = -F(r) dt^2 + \frac{1}{G(r)} dr^2 + r^2 ( d\theta^2 + \sin^2 \theta d\phi^2 ).
\end{equation}
We assume that the above metric admits a degenerate horizon at $r = r_H$. Namely, the metric functions, $F(r)$ and $G(r)$, which have only $r$-dependence, can be expanded around $r = r_H$ as
\begin{align}
\label{NHE_metric_ex}
F(r) &\equiv F_2 ( r - r_H )^2 + F_3 ( r - r_H )^3 + \cdots, \nonumber \\*
G(r) &\equiv G_2 ( r - r_H )^2 + G_3 ( r - r_H )^3 + \cdots.
\end{align}
The action \eqref{action_scalar} yields the scalar field equation
\begin{equation}
\label{scalEOM}
\nabla_\mu \nabla^\mu \Phi = \frac{1}{c_2} \left( c_1 \mathcal{G} \frac{\partial f}{\partial \eta} + c_3 \frac{\partial V}{\partial \eta} \right) \Phi,
\end{equation}
and the complex scalar field can be written in the form of \cite{lin}
\begin{equation}
\label{scalar}
\Phi \equiv \Psi (r) e^{-i\omega t},
\end{equation}
where we assume that $\Psi(r)$ is a real function and can be expanded around $r = r_H$ as
\begin{equation}
\label{NHE_scalar}
\Psi (r) = \Psi_0 + \Psi_1 (r-r_H) + \Psi_2 (r-r_H)^2 + \cdots.
\end{equation}
We substitute these expansions into the equation \eqref{scalEOM} and demand that the equation is satisfied in each order of $r-r_H$. The scalar field equation \eqref{scalEOM} can be expressed below in terms of the metric \eqref{metric} and the scalar field \eqref{scalar}.
\begin{align}
\label{Sbt_scalEOM}
&e^{-i\omega t} \Bigg\{ FG \Psi^{\prime \prime} + \frac{1}{2} \left( F^\prime G + FG^\prime + \frac{1}{r} FG \right) \Psi^\prime \\
&\qquad \qquad + \left( \omega - \frac{c_3}{c_2} \frac{\partial V}{\partial \eta} F + \frac{2c_1}{c_2 r^2} \frac{\partial f}{\partial \eta} \left[ \left( 2F^{\prime \prime} - \frac{F^{\prime 2}}{F} \right) \left( 1 - G \right) G + F^\prime G^\prime \left( 1 - 3G \right) \right] \right) \Psi \Bigg\} = 0.
\end{align}
We substitute expansions \eqref{NHE_metric_ex} and \eqref{NHE_scalar} into the above equation. Then the equation is written as
\begin{equation}
\label{Exp_scalEOM}
\omega^2 \Psi_0 + \omega^2 \Psi_1 \left( r - r_H \right) + \cdots + \left( \sum_{k=0}^{n-2} \alpha_{n,k} \Psi_k + \omega^2 \Psi_n \right) \left( r - r_H \right)^n = 0.
\end{equation}
Here, we introduce $\alpha_{n,k}$ as the coefficients for $\Psi_k$ of the $n$-th order to make the expression simple. For the last term of \eqref{Exp_scalEOM}, $\Psi_k$ terms vanish due to the results from the lower order terms. As a result, the coefficient of the $n$-th order as $n = 0,1,2,3,\cdots$ is
\begin{equation}
\label{kleingordonn_ex}
\omega^2 \Psi_n = 0.
\end{equation}
Therefore, complex scalar fields (for which $\omega \neq 0$) vanish throughout on and outside the degenerated horizon. The results for the non-degenerated horizon case \cite{lin} and degenerated horizon case are shown in table \ref{DP_scalar}.

\begin{table}[htbp]
\caption{The results of the analysis for the ``complex scalar field'' in the case of GB and R$\Lambda$GB coupling.}
\label{DP_scalar}
\centering
\begin{tabular}{l|c|c}
\toprule
								& $\Psi_i$		& Values of the coefficients			\\
\midrule
\multirow{2}{*}{Non-degenerated horizon}	& $i = 0$		& 0								\\
								& $\vdots$	&								\\
\cmidrule{1-3}
\multirow{3}{*}{Degenerated horizon}		& $i = 0$		& \multirow{3}{*}{all coefficients vanish}	\\
								& $\vdots$	&								\\
								& $i = n$		&								\\
\bottomrule
\end{tabular}
\end{table}

\section{No-go theorem for black holes with the complex vector hair}
\label{nogo_vector}

In this section, applying the same method developed in the previous section, we will show a no-go theorem that in the $\Lambda$-VTGB theory, a spherically symmetric, static black hole cannot admit a complex vector field hair. We will show this no-go theorem for both the extremal and the non-extremal black hole cases.

\subsection{Theoretical preparations}

Here we consider the vector-tensor-Gauss-Bonnet gravity with a cosmological constant $\Lambda$ described by the action
\begin{equation}
\label{action_vector}
\mathcal{S} = \int d^4x \sqrt{-g} \left[ R - 2\Lambda + c_1 h(\zeta) \mathcal{G} + c_4 F_{\mu \nu} \bar{F}^{\mu \nu} + c_5 U(\zeta) \right],
\end{equation}
where $F_{\mu \nu} \equiv \nabla_\mu X_\nu - \nabla_\nu X_\mu$ is the field strength tensor of complex vector fields $X_\mu$. $h(\zeta)$ and $U(\zeta)$ are the non-minimal coupling function and the potential which are  arbitrary functions of $\zeta \equiv X_\mu \bar{X}^\mu$. In this theory, we will establish the following theorem: {\it In general Einstein-$\Lambda$-VTGB systems, static spherically symmetric black holes cannot have an axial sector and polar sector hair of complex vector fields. Furthermore, complex vector fields vanish over the spacetime (i.e., on and outside the event horizon), when the event horizon is degenerated}.

From the action \eqref{action_vector}, we can obtain the following vector field equation
\begin{equation}
\label{vecEOM}
\nabla_\nu {F_\mu}^{\nu} = - \frac{1}{2c_4} \left( c_1 \mathcal{G} \frac{\partial h}{\partial \zeta} + c_5 \frac{\partial U}{\partial \zeta} \right) X_\mu,
\end{equation}
and Einstein equation
\begin{equation}
\label{gravEOM}
R_{\mu \nu} - \frac{1}{2} Rg_{\mu \nu} + \Lambda g_{\mu \nu} = T^{GB}_{\mu \nu} + T^X_{\mu \nu},
\end{equation}
where the energy momentum tensor for the Gauss-Bonnet term $T^{GB}_{\mu \nu}$ is described as
\begin{equation}
\label{geoEMT}
T^{GB}_{\mu \nu} \equiv - 4c_1 P_{\lambda \mu \rho \nu} ( \nabla^\lambda \nabla^\rho h ),
\end{equation}
with the tensor
\begin{equation}
\label{geometricP}
P_{\lambda \mu \rho \nu} \equiv R_{\lambda \mu \rho \nu} + g_{\lambda \nu} R_{\mu \rho} - g_{\lambda \rho} R_{\mu \nu} + g_{\mu \rho} R_{\lambda \nu} - g_{\mu \nu} R_{\lambda \rho} + \frac{R}{2} ( g_{\lambda \rho} g_{\mu \nu} - g_{\lambda \nu} g_{\mu \rho} ),
\end{equation}
and that for the vector field $T^X_{\mu \nu}$ is described as
\begin{equation}
\label{vecEMT}
T^X_{\mu \nu} \equiv - c_4 \left( 2F_{( \nu \alpha} \bar{F}_{\mu )}^{\ \ \alpha} - \frac{1}{2} g_{\mu \nu} F_{\alpha \beta} \bar{F}^{\alpha \beta} \right) - c_5 \left( \frac{\partial U}{\partial \zeta} X_\mu \bar{X}_\nu - \frac{1}{2} U g_{\mu \nu} \right).
\end{equation}
We again employ the metric ansatz \eqref{metric} as in the scalar case. Also we express complex vector fields $X_\mu$ as
\begin{align}
\label{vector}
X_\mu &= e^{-i\omega t} \left[ \left(
\begin{array}{c}
0 \\*
0 \\*
a(r) \csc \theta \partial_\phi Y(\theta ,\phi) \\*
-a(r) \sin \theta \partial_\theta Y(\theta ,\phi)
\end{array}
\right) + \left(
\begin{array}{c}
d(r) Y(\theta ,\phi) \\*
e(r) Y(\theta ,\phi) \\*
b(r) \partial_\theta Y(\theta ,\phi) \\*
b(r) \partial_\phi Y(\theta ,\phi)
\end{array}
\right) \right],
\end{align}
with spherical harmonics $Y(\theta, \phi)$. We refer to the first term of vector fields \eqref{vector} as ``axial sector'', while the second term of that as ``polar sector'' associated with their parity\cite{annulli}. We also assume that $a(r)$, $b(r)$, $d(r)$ and $e(r)$ are real functions.

When the event horizon is not degenerated, the metric functions $F(r)$ and $G(r)$ are expanded around the event horizon $r = r_H$ as
\begin{align}
\label{NHE_metric}
F(r) &\equiv F_1 ( r - r_H ) + F_2 ( r - r_H )^2 + \cdots, \nonumber \\*
G(r) &\equiv G_1 ( r - r_H ) + G_2 ( r - r_H )^2 + \cdots,
\end{align}
while the $r$-dependent coefficients of the vector fields are expanded as
\begin{align}
\label{NHE_avector}
a(r) &\equiv a_0 + a_1 ( r - r_H ) + a_2 ( r - r_H )^2 + \cdots,
\end{align}
and
\begin{align}
\label{NHE_pvector}
b(r) &\equiv b_0 + b_1 ( r - r_H ) + b_2 ( r - r_H )^2 + \cdots, \nonumber \\*
d(r) &\equiv d_0 + d_1 ( r - r_H ) + d_2 ( r - r_H )^2 + \cdots, \nonumber \\*
e(r) &\equiv e_0 + e_1 ( r - r_H ) + e_2 ( r - r_H )^2 + \cdots.
\end{align}
If the coefficients on the horizon $a_0$, $b_0$, $d_0$ and $e_0$ vanish, then we can show that black holes possess no complex vector hair. In the next section, we confirm that in detail for axial sector and polar sector separately.

\subsection{Results of the axial sector}
\label{nogo_axial}

In this section, we show the no-go theorem for the axial sector of complex vector fields \eqref{vector}. First, we substitute near-horizon expansions of the metric \eqref{NHE_metric} and the vector field \eqref{NHE_avector} into the vector field equation \eqref{vecEOM}. Then, we demand that the equation is satisfied in each order of $r-r_H$. Obtained results are shown in table \ref{DP_axial}.

\subsubsection{Non-degenerated horizon case}
\label{axial_noex}

The coefficient of the 0th order of the vector field equation for the angular component is
\begin{equation}
\label{aproca_ang0}
- \omega^2 a_0 = 0.
\end{equation}
We assume $\omega \neq 0$ so that the vector field is complex. Then the above equation \eqref{aproca_ang0} is satisfied only when $a_0 = 0$, and hence the axial sector of complex vectors must vanish on the horizon.

\subsubsection{Degenerated horizon case}
\label{axial_ex}

Next, we set $F_1 = 0 = G_1$ in the metric expansions \eqref{NHE_metric} in order to consider the extremal horizon case. We perform similar procedures of \eqref{Sbt_scalEOM}-\eqref{Exp_scalEOM}. Then the coefficient of the $n$-th order of the vector field equation for the angular component in terms of $n = 0,1,2,3,\cdots$ is
\begin{equation}
\label{aproca_angn_ex}
- \omega^2 a_n = 0.
\end{equation}
This equation is satisfied only for $a_n = 0$ as long as we consider a complex vector field. Therefore, we find that the axial sector of complex vectors must vanish not only on the degenerated horizon, but also outside the horizon.

\begin{table}[htbp]
\caption{For the axial sector, the results of the analysis for the ``complex vector field'' in the case of GB and R$\Lambda$GB coupling.}
\label{DP_axial}
\centering
\begin{tabular}{l|c|c}
\toprule
								& $a_i$		& The values of the coefficients			\\
\midrule
\multirow{2}{*}{Non-degenerated horizon}	& $i = 0$		& 0								\\
								& $\vdots$	&								\\
\cmidrule{1-3}
\multirow{3}{*}{Degenerated horizon}		& $i = 0$		& \multirow{3}{*}{all coefficients vanish}	\\
								& $\vdots$	&								\\
								& $i = n$		&								\\
\bottomrule
\end{tabular}
\end{table}

\subsection{Results of the polar sector}
\label{nogo_polar}

In this section, we substitute the near-horizon expansions \eqref{NHE_metric} and \eqref{NHE_pvector} into the vector field equation \eqref{vecEOM} and Lorenz condition $\nabla_\mu X^\mu = 0$, and examine the equations in each order of $r - r_H$ as in section \ref{axial_noex}. Obtained results are shown in table \ref{DP_polar}.

\subsubsection{Non-degenerated horizon case}
\label{polar_noex}
 
Concerning coefficients of the 0th order of the equations, the temporal and angular components of the vector field equation are
\begin{equation}
\label{pproca_t0}
\left[ \frac{2c_1}{c_4} \frac{\partial h}{\partial \zeta} G_1 \left( \frac{3F_2}{F_1} + \frac{G_2}{G_1} - 2G_1 \right) - \frac{c_5}{c_4} \frac{\partial U}{\partial \zeta} r_H^2 + 2\ell ( \ell + 1 ) \right] d_0 - 2i\omega \ell ( \ell + 1 ) b_0 = 0,
\end{equation}
\begin{equation}
\label{pproca_ang0}
2i\omega d_0 - 2\omega^2 b_0 = 0,
\end{equation}
respectively. The Lorenz condition reduces to
\begin{equation}
\label{plorenz0}
2i\omega r_H^2 d_0 = 0.
\end{equation}
The equation \eqref{plorenz0} is satisfied only when $d_0 = 0$ as long as we assume complex vectors ($\omega \neq 0$) and black holes ($r_H \neq 0$). Then we obtain $b_0 = 0$ from \eqref{pproca_ang0}. Moreover we find that \eqref{pproca_t0} is satisfied automatically when $d_0 = 0 = b_0$.

With the above results about the 0th order analysis at our hand, we next investigate the coefficients of the 1st order of $r-r_H$. The temporal, radial and angular components of the vector field equation are 
\begin{align}
\label{pproca_t1}
&4r_H^2 G_1 d_2 + \left[ \frac{2c_1}{c_4} \frac{\partial h}{\partial \zeta} G_1 \left( \frac{3F_2}{F_1} + \frac{G_2}{G_1} - 2G_1 \right) - \frac{c_5}{c_4} \frac{\partial U}{\partial \zeta} r_H^2 + 2\ell ( \ell + 1 ) + r_H^2 G_1 \left( \frac{4}{r} - \frac{F_2}{F_1} + \frac{G_2}{G_1} \right) \right] d_1 \nonumber \\
&\quad + 2i\omega r_H^2 G_1 e_1 + i\omega r_H^2 G_1 \left( \frac{4}{r} - \frac{F_2}{F_1} + \frac{G_2}{G_1} \right) e_0 - 2i\omega \ell ( \ell + 1 ) b_1 = 0,
\end{align}
\begin{equation}
\label{pproca_r1}
2G_1 \left( \frac{2c_1}{c_4} \frac{\partial h}{\partial \zeta} + \omega r_H^2 \right) e_0 + 2i\omega r_H^2 G_1 d_1 = 0,
\end{equation}
\begin{equation}
\label{pproca_ang1}
2F_1 G_1 e_0 + 2i\omega d_1 - 2 \left( F_1 G_1 + \omega^2 \right) b_1 = 0,
\end{equation}
respectively. The Lorenz condition yields
\begin{equation}
\label{plorenz1}
2r_H^2 F_1 G_1 e_0 + 2i\omega r_H^2 d_1 = 0.
\end{equation}
Lorenz condition \eqref{plorenz1} is satisfied when
\begin{equation}
\label{noex_e0}
e_0 = - \frac{i\omega}{F_1 G_1} d_1.
\end{equation}
Since the coefficients $d$ and $e$ are real, \eqref{noex_e0} is realized only when $e_0 = 0 = d_1$. Combined with the results for the 0th order, we find $d_0 = 0 = e_0 = b_0$. Thus, the polar sector of complex vectors must vanish on the horizon.

\subsubsection{Degenerated horizon case}
\label{polar_ex}

Now we consider the extremal horizon case for the polar sector in a similar way as in section \ref{axial_ex}. The temporal and angular components of the 0th order of the vector field equation are
\begin{equation}
\label{pproca_t0_ex}
\left[ \frac{2c_1}{c_4} \frac{\partial h}{\partial \zeta} G_2 - \frac{c_5}{c_4} \frac{\partial U}{\partial \zeta} r_H^2 + 2\ell ( \ell + 1 ) \right] d_0 - 2i\omega \ell ( \ell + 1 ) b_0 = 0,
\end{equation}
\begin{equation}
\label{pproca_ang0_ex}
2i\omega d_0 - 2\omega^2 b_0 = 0,
\end{equation}
respectively. Also we obtain
\begin{equation}
\label{plorenz0_ex}
2i\omega r_H^2 d_0 = 0,
\end{equation}
from the Lorenz condition. From \eqref{plorenz0_ex} and therefore \eqref{pproca_ang0_ex}, we find $d_0 = 0 = b_0$. Next, in the 1st order, the temporal and angular components of the vector field equation are
\begin{equation}
\label{pproca_t1_ex}
\left[ \frac{2c_1}{c_4} \frac{\partial h}{\partial \zeta} G_2 - \frac{c_5}{c_4} \frac{\partial U}{\partial \zeta} r_H^2 + 2\ell ( \ell + 1 ) \right] d_1 - 2i\omega \ell ( \ell + 1 ) b_1 = 0,
\end{equation}
\begin{equation}
\label{pproca_ang1_ex}
2i\omega d_1 - 2\omega^2 b_1 = 0,
\end{equation}
respectively and from Lorenz condition,
\begin{equation}
\label{plorenz1_ex}
2i\omega r_H^2 d_1 = 0,
\end{equation}
is obtained. Then $d_1 = 0 = b_1$ results from \eqref{pproca_ang1_ex} and \eqref{plorenz1_ex}. Additionally, we calculate the coefficients of the 2nd order as well. The temporal, radial and angular components are
\begin{align}
\label{pproca_t2_ex}
&\left[ \frac{8c_1}{c_4} \frac{\partial h}{\partial \zeta} G_2 - \frac{c_5}{c_4} \frac{\partial U}{\partial \zeta} r_H^2 + 2\ell ( \ell + 1 ) \right] d_2 - 2i\omega \ell ( \ell + 1 ) b_2 \nonumber \\
&\quad + 2i\omega r_H^2 G_2 e_1 + i\omega r_H^2 G_2 \left( \frac{4}{r} - \frac{F_3}{F_2} + \frac{G_3}{G_2} \right) e_0 = 0,
\end{align}
\begin{equation}
\label{pproca_r2_ex}
2\omega r_H^2 G_2 e_0 = 0,
\end{equation}
\begin{equation}
\label{pproca_ang2_ex}
2i\omega d_2 - 2\omega^2 b_2 = 0,
\end{equation}
respectively and
\begin{equation}
\label{plorenz2_ex}
2i\omega r_H^2 d_2 = 0,
\end{equation}
is derived from the Lorenz condition. Then we find $e_0 = 0$ from \eqref{pproca_r2_ex}, and $d_2 = 0 = b_2$ from \eqref{pproca_ang2_ex} and \eqref{plorenz2_ex}. As a result, \eqref{pproca_t2_ex} yields $e_1 = 0$. For the case of extremal black holes, we need to expand the equation \eqref{vecEOM} and the Lorenz condition to the 2nd order, to obtain the result $d_0 = 0 = e_0 = b_0$ so as to show that complex vector fields vanish on the degenerated horizon.

Furthermore, to confirm the behavior of the $n$-th order of the polar sector as in section \ref{axial_ex}, we calculate the coefficients of the 3rd order or more. The coefficients of the $n$-th order for the temporal and angular components of the vector field equation and the Lorenz condition are
\begin{equation}
\label{pproca_tn_ex}
\left[ \frac{8c_1}{c_4} \frac{\partial h}{\partial \zeta} G_2 - \frac{c_5}{c_4} \frac{\partial U}{\partial \zeta} r_H^2 + 2\ell ( \ell + 1 ) + 2n ( n - 1 ) r_H^2 G_2 \right] d_n - 2i\omega \ell ( \ell + 1 ) b_n + 2i\omega ( n - 1 ) r_H^2 G_2 e_{n-1} = 0,
\end{equation}
\begin{equation}
\label{pproca_angn_ex}
2\omega \left( i d_n - \omega b_n \right) = 0,
\end{equation}
\begin{equation}
\label{plorenzn_ex}
2i\omega r_H^2 d_n = 0,
\end{equation}
respectively, and these equations yield $d_n = 0 = e_{n-1} = b_n$. Therefore, it is shown that the polar sector of complex vectors in the spacetime which has a degenerated horizon vanishes not only on the horizon, but also outside the horizon.

\begin{table}[htbp]
\caption{For the polar sector, the results of the analysis for the ``complex vector field'' in the case of GB and R$\Lambda$GB coupling.}
\label{DP_polar}
\centering
\begin{tabular}{l|C{1cm}|c}
\toprule
\multicolumn{2}{c|}{}										& The values of the coefficients				\\
\midrule
\multirow{6}{*}{Non-degenerated horizon}	& \multirow{2}{*}{$d_i$}	& \multirow{6}{*}{coefficients of $i=0$ vanish}	\\
								&					&									\\
\cline{2-2}
								& \multirow{2}{*}{$e_i$}	& 									\\
								&					&									\\
\cline{2-2}
								& \multirow{2}{*}{$b_i$}	&									\\
								&					&									\\
\cmidrule{1-3}
\multirow{6}{*}{Degenerated horizon}		& \multirow{2}{*}{$d_i$}	& \multirow{6}{*}{all coefficients vanish}		\\
								&					&									\\
\cline{2-2}
								& \multirow{2}{*}{$e_i$}	&									\\
								&					&									\\
\cline{2-2}
								& \multirow{2}{*}{$b_i$}	&									\\
								&					&									\\
\bottomrule
\end{tabular}
\end{table}

\subsection{Results of the full vector}
\label{nogo_full}

We calculated only for the axial sector or the polar sector in section \ref{nogo_axial} and \ref{nogo_polar}, respectively. From now on, we analyze for the full vector \eqref{vector} possessing both the axial sector and polar sector. Obtained results are shown in table \ref{DP_full}.

\subsubsection{Non-degenerated horizon case}
\label{full_noex}

The coefficients of the 0th order of the vector field equation for radial and angular components are
\begin{equation}
\label{fproca_r0}
i d_1 - \omega e_0 = 0,
\end{equation}
\begin{equation}
\label{fproca_phi0}
\left( id_0 - \omega b_0 \right) \partial_\phi Y + \omega a_0 \sin \theta \partial_\theta Y = 0,
\end{equation}
and the coefficient of the Lorenz condition is
\begin{equation}
\label{fproca_lorenz0}
i\omega d_0 = 0.
\end{equation}
We again onsider $\omega \neq 0$, so that the vector field is complex. We, therefore, obtain $d_0 = 0$ from the equation \eqref{fproca_lorenz0}. Since the metric functions $d(r)$ and $e(r)$ are assumed to be real, the equation \eqref{fproca_r0} is satisfied only when $e_0 = 0 = d_1$. Solving the equation \eqref{fproca_phi0} for $b_0$ with $d_0 = 0$, we can get
\begin{align*}
b_0 &= a_0 \sin \theta \frac{\partial_\theta Y}{\partial_\phi Y},\\
&= a_0 \sin \theta \frac{\partial \phi}{\partial \theta}.
\end{align*}
We can immediately find $b_0 = 0$, because the coordinate $\theta$ orthogonal to $\phi$. Solving the equation \eqref{fproca_phi0} for $a_0$, we can show $a_0 = 0$ in the same argument. We, therefore, indicated that the complex vector field vanishes on the event horizon, i.e., we proved no-hair theorem for the complex vector field.

\subsubsection{Degenerated horizon case}
\label{full_ex}

Next, we deal with the case that the event horizon is degenerated. For $n = 0,1,2,\cdots$, the coefficients of the $n$-th order of the vector field equation for radial and angular components are
\begin{equation}
\label{fproca_rn_ex}
i (n+1) d_{n+1} - \omega e_0 = 0,
\end{equation}
\begin{equation}
\label{fproca_phin_ex}
\left( id_n - \omega b_n \right) \partial_\phi Y + \omega a_n \sin \theta \partial_\theta Y = 0,
\end{equation}
and the coefficient of the Lorenz condition is
\begin{equation}
\label{fproca_lorenz_ex}
i\omega d_n = 0.
\end{equation}
We obtained the results $a_n = 0 = b_n = d_n = e_n$ in the same analysis as the case of non-degenerated horizon. We conclude that the complex vector field vanishes on and outside the horizon when the event horizon is degenerated.

\begin{table}[htbp]
\caption{For the full vector, the tabulated results of the analysis for the ``complex vector field'' in the case of GB and R$\Lambda$GB coupling.}
\label{DP_full}
\centering
\begin{tabular}{l|C{1cm}|c}
\toprule
\multicolumn{2}{c|}{}										& The values of the coefficients				\\
\midrule
\multirow{8}{*}{Non-degenerated horizon}	& \multirow{2}{*}{$a_i$}	& \multirow{8}{*}{coefficients of $i=0$ vanish}	\\
								&					&									\\
\cline{2-2}
								& \multirow{2}{*}{$d_i$}	&									\\
								&					&									\\
\cline{2-2}
								& \multirow{2}{*}{$e_i$}	&									\\
								&					&									\\
\cline{2-2}
								& \multirow{2}{*}{$b_i$}	&									\\
								&					&									\\
\cmidrule{1-3}
\multirow{8}{*}{Degenerated horizon}		& \multirow{2}{*}{$a_i$}	& \multirow{8}{*}{all coefficients vanish}		\\
								&					&									\\
\cline{2-2}
								& \multirow{2}{*}{$d_i$}	&									\\
								&					&									\\
\cline{2-2}
								& \multirow{2}{*}{$e_i$}	&									\\
								&					&									\\
\cline{2-2}
								& \multirow{2}{*}{$b_i$}	&									\\
								&					&									\\
\bottomrule
\end{tabular}
\end{table}

\section{Extra coupled terms and coupling dependence of the no-hair theorem}
\label{EC_CD}

\subsection{Non-minimal coupling to the Ricci scalar and the cosmological constant}
\label{extra_source}

In the previous sections, the coupling function is multiplied only by the GB invariant (from now on, we name this case "GB coupling"). In this section, we will consider the theory written as
\begin{equation}
\label{action_maxsource}
\mathcal{S} = \int d^4x \sqrt{-g} \left[ f(\eta) \left( R - 2\Lambda + c_1 \mathcal{G} \right) + \mathcal{L}_{\rm matter} \right],
\end{equation}
where the coupling function is also multiplied by the Ricci scalar and the cosmological constant (then we name this case ``R$\Lambda$GB coupling''). We obtain apparently the same result with the GB coupling case shown in tables \ref{DP_scalar}-\ref{DP_full}. 

Considering an extra coupling to the Ricci scalar and the cosmological constant means adding an extra term that includes the Ricci scalar and the cosmological constant as a source, to the effective mass term in the field equation. Here, an effective mass term plays a role of damping. Then, an increasing of the effective mass causes a stronger damping of the matter field, and therefore the no-hair is boosted. For this reason, increasing non-minimally coupled terms cannot change the no-hair theorem.

\subsection{Coupling dependence of the no-go theorem}
\label{coupling_dependence}

In the previous sections, we consider $\Phi \Phi^\ast = | \Phi |^2$ for the complex scalar field and $X_\mu \bar{X}^\mu = | X |^2$ for the complex vector field as the argument of the coupling function and the field potential. In this section, we will modify our argument to examine the coupling dependence of the no-hair theorem. For example, in the case of scalar fields, we suppose $| \Phi |^{2p}$ with $p$ some real parameter as the argument and calculate for the different value of $p$. Of course, $p = 1$ corresponds to the case of \S \ref{nogo_scalar} - \ref{nogo_vector}. Hereafter, in addition to $p = 1$ (quadratic) case, we show the results for $p = \frac{1}{2}$ (linear) case and $p = -\frac{1}{2}, -1$ (fractional) case. As correspondents of $p = 0$ case, we adopt $\ln | \Phi |^2$ and $\ln | X |^2$, and name them logarithmic case.

Since the results in the non-degenerated horizon case have no significant differences, we do not list them in the following tables \ref{CD_scalar}-\ref{CD_full}. For the complex scalar, we show the results in table \ref{CD_scalar}. We can find that the complex scalar field vanishes on and outside the horizon only for $p = 1$ in degenerated horizon case. For the complex vector, we show the results in table \ref{CD_axial} for axial sector and polar sector. The results of axial sector is qualitatively the same as that of the scalar field. For the polar sector, we obtain different results. In the case of non-degenerated horizon, we show that no-hair holds for all investigated values of $p$. While in the case of degenerated horizon, we prove that the temporal and angular components vanish on the horizon. However the radial component does not vanish on the horizon and possesses a possibility of hair. As well as the case of the scalar field and axial vector field, we prove that the vector field vanishes on and outside the horizon only for the case of $p = 1$ and degenerated horizon. Finally, we present the results of the full vector field in table \ref{CD_full}. As well as the case of the axial sector and the polar sector, it is found that the couplings of $p \neq 1$ for the degenerated horizon case yield a possibility of a hairy black hole. For the complex vector, we conclude as follows: In the case of non-degenerated horizon, we proved no-hair theorem for all examined values of $p$. While in the case of degenerated horizon, the possibility of the radial vector hair exists on the horizon and only temporal component vanishes on and outside the horizon for $p \neq 1$.

The lower a value of $p$ gets, the weaker the strength of a coupling between a matter field and a source term becomes. In terms of the field equation, the damping effect of the source term in the equation becomes smaller. Therefore, the vector field is easy to exist in the spacetime. In the case of degenerated horizon, indeed, both of the scalar field and the vector field do not vanish all over the spacetime for $p \neq 0$. we note that a physical interpretation about the conditions in which the radial functions of matter field become non-zero, is still uninvestigated.

\begin{table}[H]
\caption{The tabulated results of the analysis for ``the complex scalar field'' with different values of $p$ in degenerated horizon case.}
\label{CD_scalar}
\centering
\begin{tabular}{cc|C{2cm}C{2.5cm}}
\toprule
\multicolumn{2}{r|}{$p =$}				& $1$	& $\frac{1}{2},\ 0,\ -\frac{1}{2},\ -1$	\\
\midrule
\multirow{3}{*}{$\Psi_i$}	& $i = 0$		& \multicolumn{2}{c}{0}					\\
					& $\vdots$	& \multicolumn{2}{c}{}						\\
					& $i = n$		& 0		& vanish or finite				\\
\bottomrule
\end{tabular}
\end{table}

\begin{table}[H]
\caption{For the axial and polar sector, the tabulated results of the analysis for ``the complex vector field'' with different values of $p$ in degenerated horizon case.}
\label{CD_axial}
\centering
\begin{tabular}{l|cc|C{2cm}C{2.5cm}}
\toprule
\multicolumn{3}{r|}{$p =$}										& $1$	& $\frac{1}{2},\ 0,\ -\frac{1}{2},\ -1$	\\
\midrule
\multirow{3}{*}{Axial sector}	& \multirow{3}{*}{$a_i$}	& $i = 0$		& \multicolumn{2}{c}{0}					\\
						&					& $\vdots$	& \multicolumn{2}{c}{}						\\
						&					& $i = n$		& 0		& vanish or finite				\\
\cmidrule{1-5}
\multirow{9}{*}{Polar sector}	& \multirow{3}{*}{$d_i$}	& $i = 0$		& \multicolumn{2}{c}{0}					\\
						&					& $\vdots$	& \multicolumn{2}{c}{}						\\
						&					& $i = n$		& \multicolumn{2}{c}{0}					\\
\cline{2-5}
						& \multirow{3}{*}{$e_i$}	& $i = 0$		& 0		& vanish or finite				\\
						&					& $\vdots$	& \multicolumn{2}{c}{}						\\
						&					& $i = 1$		& 0		& vanish or finite				\\
\cline{2-5}
						& \multirow{3}{*}{$b_i$}	& $i = 0$		& \multicolumn{2}{c}{0}					\\
						&					& $\vdots$	& \multicolumn{2}{c}{}						\\
						&					& $i = 1$		& 0		& vanish or finite				\\
\bottomrule
\end{tabular}
\end{table}

\begin{table}[H]
\caption{For the full vector, the tabulated results of the analysis for ``the complex vector field'' with different values of $p$ in degenerated horizon case.}
\label{CD_full}
\centering
\begin{tabular}{cc|C{2cm}C{2.5cm}}
\toprule
\multicolumn{2}{r|}{$p =$}				& $1$	& $\frac{1}{2},\ 0,\ -\frac{1}{2},\ -1$	\\
\midrule
\multirow{3}{*}{$a_i$}		& $i = 0$		& \multicolumn{2}{c}{0} 					\\
					& $\vdots$	& \multicolumn{2}{c}{}						\\
					& $i = n$		& 0		& vanish or finite				\\
\cline{1-4}
\multirow{3}{*}{$d_i$}		& $i = 0$		& \multicolumn{2}{c}{0}					\\
					& $\vdots$	& \multicolumn{2}{c}{}						\\
					& $i = n$		& \multicolumn{2}{c}{0}					\\
\cline{1-4}
\multirow{3}{*}{$e_i$}		& $i = 0$		& 0		& vanish or finite				\\
					& $\vdots$	& \multicolumn{2}{c}{}						\\
					& $i = 1$		& 0		& vanish or finite				\\
\cline{1-4}
\multirow{3}{*}{$b_i$}		& $i = 0$		& \multicolumn{2}{c}{0}					\\
					& $\vdots$	& \multicolumn{2}{c}{}						\\
					& $i = 1$		& 0		& vanish or finite				\\
\bottomrule
\end{tabular}
\end{table}

\section{Conclusions}
\label{concl}

In this paper, we established the following theorem for complex scalar fields: {\it In general Einstein-$\Lambda$-STGB systems, static spherically symmetric black holes cannot have a complex scalar hair. Furthermore, complex scalar fields vanish over the spacetime (i.e., on and outside the event horizon), when the event horizon is degenerated}. We have also established the theorem for complex vector fields: {\it In general Einstein-$\Lambda$-VTGB systems, static spherically symmetric black holes cannot have an axial sector and/or polar sector hair of complex vector fields. Furthermore, complex vector fields vanish over the spacetime (i.e., on and outside the event horizon), when the event horizon is degenerated}. In addition, we confirmed that the no-hair theorem is satisfied for the case in which the matter field non-minimally couples with Ricci scalar and cosmological constant in addition to GB invariant. It was also found that the non-extremal black hole has no hair in the theory with different coupling functions, while extremal black hole may possibly admit radial vector hair.

For future work, it is interesting to apply the method used in this paper to show similar type of no-go theorems for hairy black holes in other alternative gravity theories or Einstein gravity with other matter fields, e.g., tensor or Dirac fields. It would also be interesting to consider a generalization of the present results to the rotating black hole case.

\section*{Acknowledgment}
We thank Akihiro Ishibashi for helpful comments and instructions on the draft. We also thank all members of our laboratory for discussions.
The work was supported in part by JSPS KAKENHI Grants No. 21H05182, 21H05186, 20K03938, 15K05092.

\hrulefill

\end{document}